# Ultrasound Classification of Breast Masses Using a Comprehensive Nakagami Imaging and Machine Learning Framework

Ahmad Chowdhury [a], Rezwana R. Razzaque [a], Ahmad Shafiullah [a], Sabiq Muhtadi [a], Brian S. Garra [b] and S. Kaisar Alam [c, d]

[a] Department of Electrical and Electronic Engineering, Islamic University of Technology, Gazipur, Bangladesh; [b] Division of Imaging, Diagnostics and Software Reliability, Office of Science and Engineering Laboratories, Center for Devices and Radiological Health, Food and Drug Administration, Silver Spring, MD, United States; [c] Imagine Consulting LLC, Dayton, NJ, United States; [d] The Center for Computational Biomedicine Imaging and Modelling (CBIM), Rutgers University, NJ, Piscataway, United States;

**Abstract—** In this study we investigate the potential of parametric images formed from ultrasound B-mode scans using the Nakagami distribution for non-invasive classification of breast lesions. Through a sliding window technique, we generated seven types of parametric images from each patient scan in our dataset using basic and as well as derived parameters of the Nakagami distribution. To determine the most suitable window size for image generation, we conducted an empirical analysis using three windows, and selected the best one for our study. From the parametric images formed for each patient, we extracted a total of 72 features. Feature selection was performed to find the optimum subset of features for the best classification performance. Incorporating the selected subset of features with the Support Vector Machine (SVM) classifier, and by tuning the decision threshold, we obtained a maximum classification accuracy of 93.08%, an Area under the ROC Curve (AUC) of 0.9712, a False Negative Rate of 0%, and a very low False Positive Rate of 8.65%. Our results indicate that the high accuracy of such a procedure may assist in the diagnostic process associated with detection of breast cancer, as well as help to reduce false positive diagnosis.

# INTRODUCTION

Breast cancer is the most commonly diagnosed cancer in females around the globe by a significant margin, with an estimated 2.3 million new cases of breast cancer being reported in 2020 (Sung et al., 2021). It also accounted for 6.9% (approximately 700,000) of all cancer-related deaths worldwide (Sung et al., 2021). In the United States, about 1 in 8 women (13%) are expected to develop invasive breast cancer over the course of their lifetime, and a staggering 43,600 deaths are expected in 2021 alone (American Cancer Society, 2021). The survival of patients is fundamentally associated with rapid and accurate early diagnosis, as it plays a vital role in treatment selection as well as predicting the patient's response to therapy.

Improvements in the quality of biomedical imaging have resulted in increased sensitivity to potentially harmful tissue features, such as cancer. However, this improvement in sensitivity has not been accompanied by improved specificity, or the ability to determine whether a cancer is benign or malignant. As a result, an 'overdiagnosis' problem has occurred, which is leading to 'overtreatment' (Esserman et al., 2014; Welch and Black, 2010), and current screening methods for breast cancer are associated with a high rate of false positives. In the United States, false positive recalls and screenings associated with breast cancer cost around $4 billion a year (Vlahiotis et al., 2018). This not only amounts to excessive anxiety for the patient, but can also lead to unnecessary biopsy (Vlahiotis et al., 2018), which is still considered to be the gold standard technique for pathological confirmation of malignancy and characterization of tumor grade (Tadayyon et al., 2014). According to the National Cancer Institute, breast biopsies in the United States have a false-positive rate of up to 71% (Mayo et al., 2019), resulting in an annual expense of $2.18 billion in biopsy procedures, which can be avoided with more accurate prior diagnosis. There is an inherent need to reduce false positive diagnosis associated with breast cancer treatment.

X-ray Mammography is the recommended screening technique associated with breast cancer, and has helped reduce mortality by upto 45% (Duffy et al., 2002). However, according to reports, patients with dense breasts have an increased likelihood of receiving a False Negative result for lesion detection using mammography (Oelze, 2012), with sensitivity for this patient group reported to be as low as 48% (Kolb et al., 2002). Ultrasound (US) imaging techniques have been principally applied in a supporting role for the diagnosis of breast cancer (Sainsbury, 2013). It has since become an important tool in the non-invasive diagnosis of breast cancer (Kolb et al., 2002; Sehgal et al., 2006), and several studies have shown that diagnostic accuracy of ultrasound approaches or exceeds that of mammography (Flobbe et al., 2003; Kolb et al., 2002; Leconte et al., 2003).

Conventional Ultrasound (US) imaging procedures are qualitative in nature, and involve assessment of tumor features according to the Breast Imaging Reporting and Data System (BI-RADS) criteria developed by the American College of Radiology (ACR) (Mendelson et al., 2013). This is reported to be faster, cheaper and more accurate than mammography (Förnvik et al., 2010; Shoma et al., 2006). However, ultrasound is inherently operator and device-dependent (Berg et al., 2012; Brem et al., 2015), which poses limitations on reproducibility.

On the other hand, Quantitative Ultrasound (QUS) methods are highly reproducible, and independent of device and operator related factors (Boote et al., 1988; Nam et al., 2012; Yao et al., 1990). QUS techniques allow for the evaluation of tissue in terms of structural and mechanical properties, and provide numerical data related to tissue features, which have the potential to improve the specificity of biomedical imaging techniques (Mamou and Oelze, 2013; Oelze and Mamou, 2016). The use of quantitative parameters allows clinicians to reduce the number of biopsies, whilst maintaining the same level of detection in terms of malignant cases (Trop et al., 2015). Several different quantitative parameters have been explored by researchers with regards to characterization of breast tissue. This includes textural parameters (Klimonda et al., 2019; Sadeghi-Naini et al., 2017), parameters related to entropy (Tsui, 2015; Tsui et al., 2017), and parameters derived from the power spectrum (Alam et al., 2011; Sadeghi-Naini et al., 2017) as well as related to the backscatter co-efficient (Nam et al., 2013; Oelze and Mamou, 2016).

The statistics of the envelope signal obtained from ultrasound scanners may also be modelled as a probability density function (PDF) in order to quantitatively analyze the scattering properties of soft tissue. Several distributions are utilized in this regard to model scattered signals in soft tissue (Destrempes and Cloutier, 2010). Among them the Nakagami distribution and the Homodyned K distribution have been frequently utilized by researchers to model scattered signals in the breast.

The homodyned K model for ultrasound echoes was initially proposed by Dutt and Greenleaf (Dutt and Greenleaf, 1994), and later modified by Hruska (Hruska, 2009), and Hruska and Oelze (Hruska and Oelze, 2009). The homodyned K parameter along with BI-RADS classifiers was later applied by Trop et al. (Trop et al., 2015), who demonstrated their ability to reduce the number of biopsies performed on breast cancer patients by 25%, whilst maintaining 100% sensitivity. Byra et al. (Byra et al., 2016) utilized parameters extracted from segmented parametric maps of homodyned K parameters in order to classify breast tumors, and obtained a maximum AUC of 0.84.

The Nakagami distribution is a simple and generalized model proposed for ultrasonic backscatter by Shankar (Mohana Shankar, 2000), and is one of the most important statistical distributions for modelling soft tissue (Nakagami, 1960). The Nakagami distribution is very popular in QUS due to its reduced computational complexity and its ability to adequately describe tissue scattering statistics. Many research groups have attempted to utilize characteristics of the Nakagami distribution in order to analyze ultrasound backscattered echo from the breast. Shankar et al. (Shankar et al., 2001) utilized two Nakagami parameters, m and α, to classify breast masses from a total of 52 patients (14 malignant, 38 benign), and reported AUC values of 0.79±0.11 and 0.828±0.10 respectively for the two parameters individually. In an attempt to enhance classification performance, Shankar, et al.(Shankar et al., 2002) proposed the concept of compounding the Nakagami parameter m, and observed an AUC of 0.8316 using the proposed technique. In another study, Shankar, et al. (Shankar et al., 2003) performed a five-parameter analysis for 99 patients (29 malignant, 70 benign) based on parameters derived from both Nakagami and K distribution at the site, boundary, spiculated region and shadow of the breast mass, and a combination of these features led to an AUC of 0.96 ±0.02. Tsui et al. (Tsui et al., 2009) found that traditional Nakagami parameters extracted with an unfocused transducer cannot be utilized to distinguish between different scatterer qualities. They proposed a novel approach for tissue characterization using a non-focused transducer termed noise-assisted Nakagami parameter based on empirical mode decomposition of noisy backscattered echoes, and proved its feasibility for ultrasound tissue characterization. In a later study, Tsui et al. (Tsui et al., 2010) used backscattered signals from breast scans to form Nakagami parametric images in order to classify breast lesions for 100 patients (50 malignant, 50 benign) and obtained an AUC of $0.81 \pm 0.04$ along with an accuracy of 82%. Liao et al. (Liao et al., 2011) performed 2-D analysis by integrating Nakagami parametric images and four texture feature parameters from B-scan images, and obtained an accuracy of 88.2% for 130 breast cancer patients. Liao et al. (Liao et al., 2012) also introduced a novel method to classify breast lesions by incorporating elasticity properties obtained from the strain-compounding imaging method, based on the acquisition of multiple frames under various strain conditions with the scatterer characteristics of Nakagami images. They conducted their study on a total of 50 patients and reported an AUC of 0.92. Dobruch-Sobczak, et al. (Dobruch-Sobczak et al., 2017) combined the BI-RADS assessment with Nakagami statistics and found it to be the most appropriate method to distinguish between benign and malignant breast lesions of BI-RADS category 4a. According to their study, in comparison to BI-RADS alone, using B-mode and the mLmin parameter from Nakagami image together gave an AUC of 0.978. Byra, et al. (Byra et al., 2017) utilized Nakagami parametric images from 458 RF data matrices in order to train a convolutional neural network with 5-fold cross-validation to classify breast lesions, and obtained an ROC area of 0.912. Hsu, et al. (Hsu et al., 2019) utilized morphological and texture features from B-scan images and Nakagami parametric images to characterize breast lesions, and reported an AUC of 0.96 and accuracy of 89.4%. Although quantitative parametric studies are conducted in the area within the lesion in most cases, Klimonda, et al. (Klimonda et al., 2018) proved that the two parameters, weighted entropy and the Nakagami shape parameter (m), calculated from the surrounding tissue of the lesion better classifies the lesion than values from within the lesion (AUC values were on average 18% higher for the surrounding tissue than within the tissue). In another study, Klimonda, et al. (Klimonda et al., 2019) proposed a multi-parametric approach by combining Nakagami shape parameter, entropy and texture parameters. They studied these parameters for both inside the lesion and the tissue surrounding the lesion, and obtained AUC of 0.92 for the surrounding tissue and 0.94 from a combination of both of them.

The paper proposes a novel framework for classification of breast masses using numerous types of Nakagami parametric images. Seven different parametric images are generated based on Nakagami alpha, omega and mu parameters from ultrasound radio-frequency (RF) data. Different types of morphometric, elemental and hybrid features are extracted from each of the parametric images, some of which are excellent for classification when grouped together. To our knowledge, no other research work has been conducted in such a comprehensive manner on Nakagami parameters and their features. While our research proposes a large number of features, it is also quite an impractical approach to incorporate all the features for classification. In a clinical scenario, generation of all the features for each patient would be quite cumbersome, and incorporation of a high number of features would increase the computation time of the classification algorithm. Thus, our framework further broadens in prospect by adopting a technique for feature selection which provides us with an optimum feature subset for the best classification performance. Incorporating the selected subset of features with the Support Vector Machine (SVM) classifier, and by tuning the decision threshold, we obtained a False Negative Rate of 0%, along with a False Positive Rate as low as 8.65%, and a maximum classification accuracy of 93.08%.

## METHODOLOGY

This study analyzes features derived from Nakagami parametric images. The process begins with the generation of envelope images from patient ultrasound RF data. 7 different Nakagami parametric images are then generated from each envelope image using 3 different window sizes. A total of 72 features are extracted from these Nakagami parametric images. Feature selection is performed using the RFE-CV algorithm in order to select the subset of features that would result in the best classification, and to eliminate redundant features. The selected subset of features is then utilized for classification purposes. Figure 1 depicts the proposed framework.

**Description of the Dataset**

The dataset utilized for this study is from ATL's PMA study (IRB approved) undertaken in 1994 (Alam et al., 2011). The dataset consists of RF data comprises biopsy proven analysis results of 130 patients, of which 104 are benign and 26 are malignant. The data were acquired at three clinical sites (Thomas Jefferson University, University of Cincinnati, and Yale University), at the time of routine ultrasonic examinations of patients scheduled for biopsy. Informed consent was obtained from each patient participating in the study. Participating patients had mammographically-visible lesions, that were either palpable or non-palpable, and which were discovered through screening, physical examination or both methods. Exclusion factors for patients consisted of: age (less than 18 years, due to legal consent limitations), prior breast carcinoma, biopsy or mastectomy, breast implant, simple cyst, pregnancy, microcalcifications not associated with a mass on sonography, and males or transsexuals. Before performing biopsy, the masses were examined in a supine position by an experienced radiologist or sonographer, using a Philips (ATL at the time) Ultrasound (Bothell, WA) UM-9 HDI scanner, with an L10-5 (7.5-MHz nominal frequency) linear-array transducer, during which standard ultrasonic breast examination procedures were employed. Multiple views were selected for every lesion by the radiologist, which included at least a radial and an anti-radial view, and in some cases additional views. Some patients had multiple masses and each mass was assigned a number, and the radiologist marked the images with the mass number. The file name convention also included the mass number.

In order to define the vascularity of each mass compared to that of the mirror image location in the opposite breast, color-flow and pulse-Doppler examination were performed, although the study described in this paper excludes the Doppler results, as these data were not saved. The mammographer scored each mammogram using a level of suspicion (LOS) on a scale of 1 to 5, where 1 denotes benign, 2 probably benign, 3 indeterminate, 4 probably malignant and 5 definitely malignant.

The RF echo-signal data was digitally obtained by interfacing a Spectrasonics Inc. (King of Prussia, PA) acquisition module with the scanner to allow acquisition of RF data. (The UM-9 HDI scanner didn't have an RF data output.) The L10-5 transducer was used at a default power level and a single transmit focal length, as selected by the operator. The RF data was sampled at 20 MHz, with an effective dynamic range of 14 bits. The acquired data was corrected for time-gain-control (TGC) before processing; TGC data was acquired before every scan.

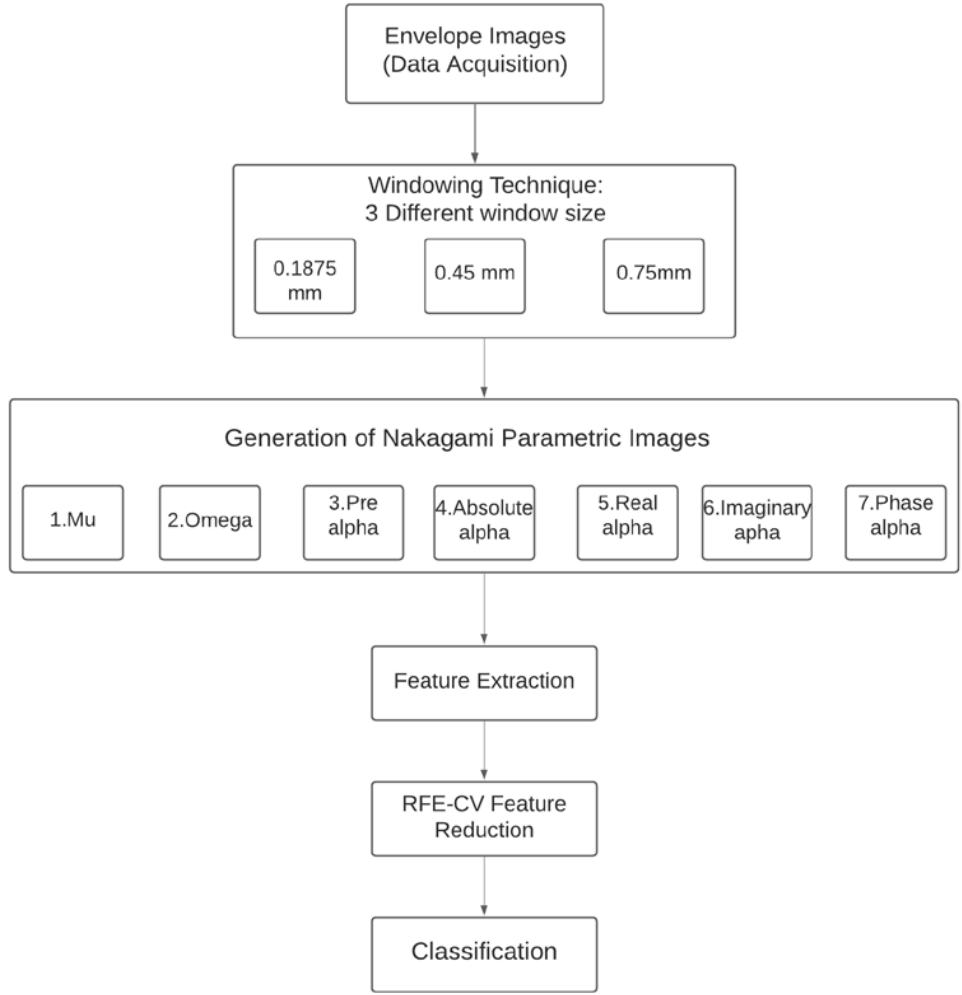

Fig. 1: Proposed Framework used in this study

**Nakagami Parametric Imaging**

*The Nakagami Distribution*

It is possible to model the backscattered echo of acoustic pulse as the algebraic sum of the contributions of individual scatterers with the tissue (Shankar, 1995; Wagner et al., 1983). Consider a total of M scatterers in the target cell, and let $X_m$ and $\varphi_m$ represent the amplitude and phase of the $m^{th}$ scatterer respectively. Then, the backscattered echo may be represented as

$$B(t) = \sum_{m=1}^{M} X_m \cos(\omega t - \varphi_m) \tag{1}$$

The backscattered echo can also be represented in terms of the in-phase and quadrature components.

$$B(t) = P\cos(\omega t) + Q\sin(\omega t) \tag{2}$$

Where,

$$P = \sum_{m=1}^{M} X_m \cos(\varphi_m) \tag{3}$$

$$Q = \sum_{m=1}^{M} X_m \sin(\varphi_m) \tag{4}$$

The envelope of the backscattered echo may be represented as

$$E = \sqrt{P^2 + Q^2} \tag{5}$$

$P$ and $Q$ may be considered Gaussian distributed with zero mean and equal variance if the target cell contains a large number of randomly located scatterers. Under this condition, the envelope $E$ will obey Raleigh statistics. However, this will not be the case if the target cell contains scatterers with randomly varying scattering cross-sections and a comparatively high degree of variance (Molthen et al., 1995; Narayanan et al., 1994; Shankar, 1995; Shankar et al., 2001). The envelope can also be considered Rician or Post Rayleigh if in addition to randomly located scatterers, the target cell contains periodically located scatterers spaced corresponding to integral multiples of wavelength at demodulation frequency (Wagner et al., 1987). All these scattering conditions exist in the Nakagami distribution (Nakagami, 1960).

The Nakagami distribution (Nakagami, 1960) is one of the simplest models for modelling backscatter envelope. It is a two-parameter distribution first introduced by Nakagami (1943, 1960) in the context of wave propagation. The Nakagami probability density function is given by,

$$N(m, \Omega) = \frac{2m^m}{\Gamma m \Omega^m} A^{2m-1} e^{-\frac{mA^2}{\Omega}} \tag{6}$$

Here, $\Gamma$ represents the Euler gamma function.

The cumulative distribution of the Nakagami distributed envelope $F(r)$ is given by,

$$F(r) = \int_{0}^{r} \frac{2m^m}{\Gamma m \Omega^m} A^{2m-1} e^{-\frac{mA^2}{\Omega}} dA \tag{7}$$

The following expressions are for the mean intensity of the Nakagami distribution and its SNR:

$$E[I] = \Omega \tag{8}$$

$$SNR^2 = m \tag{9}$$

This intensity SNR parameter should not be confused with the amplitude SNR.

The Nakagami distribution has two parameters, which are expressed as

$$m = \frac{[E(R^2)]^2}{E[R^2 - E(R^2)]^2} \tag{10}$$

$$\Omega = E[R^2] \tag{11}$$

The parameter $m$ is referred to as the shape parameter, and contains information about the envelope statistics. In the case of the Nakagami distribution, it is constrained such that $m \geq 0.5$ (Nakagami, 1960), in which case it is referred to as the Nakagami parameter. The case of $m = 1$ corresponds to Rayleigh scattering (Shankar et al., 2001), while $m$ in the range of 0.5 to 1 corresponds to pre-Rayleigh scattering (Mohana Shankar, 2000; Nakagami, 1960; Shankar, 1995). A case of $m > 1$ corresponds to post-Rayleigh or Rician (Mohana Shankar, 2000). Hence, the Nakagami distribution incorporates all these scattering conditions(Shankar et al., 2001).

The parameter $\Omega$ is a scaling parameter.

A similarity also exists between the Nakagami parameter $m$ and the effective number $M$ of the K distribution (Narayanan et al., 1994; Shankar, 1995). The K distribution was initially proposed to account for the high degree of variability in scattering cross sections of the scatterers in the target cell (Dutt and Greenleaf, 1994; Shankar, 1995).

The cumulative distribution of the K distributed envelope $F_K(r)$ can be expressed as

$$F_K(r) = \frac{2b}{\Gamma(M)} \left(\frac{br}{2}\right)^M K_{M-1}(br) \qquad r \geq 0 \quad M \geq 0 \tag{12}$$

Here, the parameter $b$ is a scaling factor, and $M$ represents the effective number of scatterers in the target cell.

There is evidence that the K distribution can be expressed as the product of two Nakagami-distributed random variables (Nakagami, 1960). Hence, the parameters of the K distribution, namely $M$ and $b$ may be expressed in terms of the parameters of the Nakagami distribution (Shankar et al., 2001) as follows

$$M = \frac{2m}{1-m} \tag{13}$$

And

$$b = 2\sqrt{\frac{2m}{\Omega(1-m)}} \tag{14}$$

Hence, the Nakagami parameter $m$ is similar to the K distribution parameter $M$, and may provide an indication of the effective number of scatterers in the target cell with a compressed dynamic range (Shankar et al., 2001).

We may define a new parameter, $\alpha$, where $\alpha = \frac{1}{b}$ and is a measure of effective cross section of scatterers in the tissue under examination. Thus,

$$\alpha = \frac{1}{2}\sqrt{\frac{\Omega(1-m)}{2m}} \tag{15}$$

It is defined as the effective cross section because of the dependance of $\alpha$ on the number density of scatterers through $m$, the signal-to-noise ratio of the cross sections through $m$, and the attenuation through both $\Omega$ or $\alpha$ (Narayanan et al., 1994; Shankar, 1995; Shankar et al., 1996). Furthermore, $\alpha$ can provide information about the scattering characteristics within the target cell through $\Omega$. Hence, the Nakagami distribution is expected to be effective in tissue characterization, as the parameters $m$ and $\alpha$ provide information about the number density of scatterers, the level of attenuation present and the degree of homogeneity in the scattering cross-sections (Shankar et al., 2001).

*Generation of Nakagami Parametric Images*

Tsui and Chang (Tsui and Chang, 2007) found that Nakagami images constructed with an optimal window size have the ability to impart both the global and local backscattered statistics of the ultrasonic signals in a tissue, leading it to be a very efficient local scatterer concentration detector. They found that the optimal window for generating Nakagami images is a square with side length equal to three times the pulse length of the incident ultrasound beam, and concluded that Nakagami images can effectively assist B-mode image in medical diagnosis (Tsui and Chang, 2007).

Usually, a smaller window size is chosen for the parametric imaging to achieve a good spatial resolution of the image, but in that case most estimators become unstable and the tissue property map becomes inaccurate (Larrue and Noble, 2011). Again, larger window size is chosen to obtain a smooth and accurate map of tissue properties, decreasing spatial resolution of the map i.e., making smaller structures and lesion boundaries undetectable (Larrue and Noble, 2011). Especially, the Nakagami parameter $m$ is sensitive to such problems of estimation, window size and boundary effects (Larrue and Noble, 2011). To solve these problems, a trade-off between the resolution and the stability of the estimator i.e., the smoothness of the parametric image has to be maintained (Larrue and Noble, 2011). Though our study focuses on automated quantitative analysis (resolution of the image is inconsequential), we tried to maintain the trade-off between the aforementioned quantities and choose an optimum window size through empirical analysis.

A Nakagami Image is the local map of Nakagami parameters generated from the envelope image. The following steps were followed to generate Nakagami images:

1) Local backscattered envelopes are collected using a window within the envelope image. The local backscattered envelope values are used to estimate local Nakagami parameters, $m$ and $\Omega$, which are allocated as new pixels in the corresponding m and omega matrices.

2) Step 1 is repeated with the window moving in 0.0385mm increments through the entire envelope image, yielding the Nakagami images as maps of local $m$ and $\Omega$ values.

Derived Nakagami images, namely Pre-alpha, Real alpha, Imaginary alpha, Phase alpha and Absolute alpha are generated from the $m$ and $\Omega$ images by estimating the derived Nakagami values corresponding to each $m$ and $\Omega$ value.

In order to conduct empirical analysis, three different datasets are generated using three different window sizes: $0.1875\ mm$, $0.45\ mm$ and $0.75mm$. For these 3 different window sizes an overlap of 80% for $0.1875\ mm$, 91.67% for $0.45\ mm$ and 95% for $0.75mm$ was maintained in step 2. Empirical analysis was conducted to determine the best size of the window for classification by observing accuracy and area under curve (AUC) for the varying window sizes.

*Description of Nakagami Images*

The parameters that have been used in this study for analysis were derived from two different types of Nakagami parametric images:

**1)** Nakagami Basic Images

**2)** Nakagami Derived Images

Nakagami Basic Images:

1. $m$ Image: map of the Nakagami parameter $m$ which provides information about envelope statistics as well as effective number of scatterers (Shankar et al., 2001). Its value is constricted such that,

$$m \geq 0.5$$

2. $\Omega$ Image: map of the parameter $\Omega$ which is a Scaling parameter.

Nakagami Derived Images:

The derived Nakagami images were formed using the parameter $\alpha$, which is a measure of effective cross-section of scatterers. It is given by:

$$\alpha = \frac{1}{2}\sqrt{\frac{\Omega(1-m)}{2m}} \qquad (15)$$

It is a complex value. 5 types of images are formed using the α parameter:

1. Pre-$\alpha$ Image: map of the parameter pre-alpha which we define as follows:

$$Pre - alpha = \frac{\Omega(1-m)}{2m} \qquad (16)$$

2. $\alpha$-absolute Image: map of the absolute value of the $\alpha$ parameter.

3. $\alpha$-phase Image: map of the phase value of the $\alpha$ parameter.

4. $\alpha$-real Image: map of the real part of the $\alpha$ parameter.

5. $\alpha$-imaginary Image: map of the imaginary part of the $\alpha$ parameter.

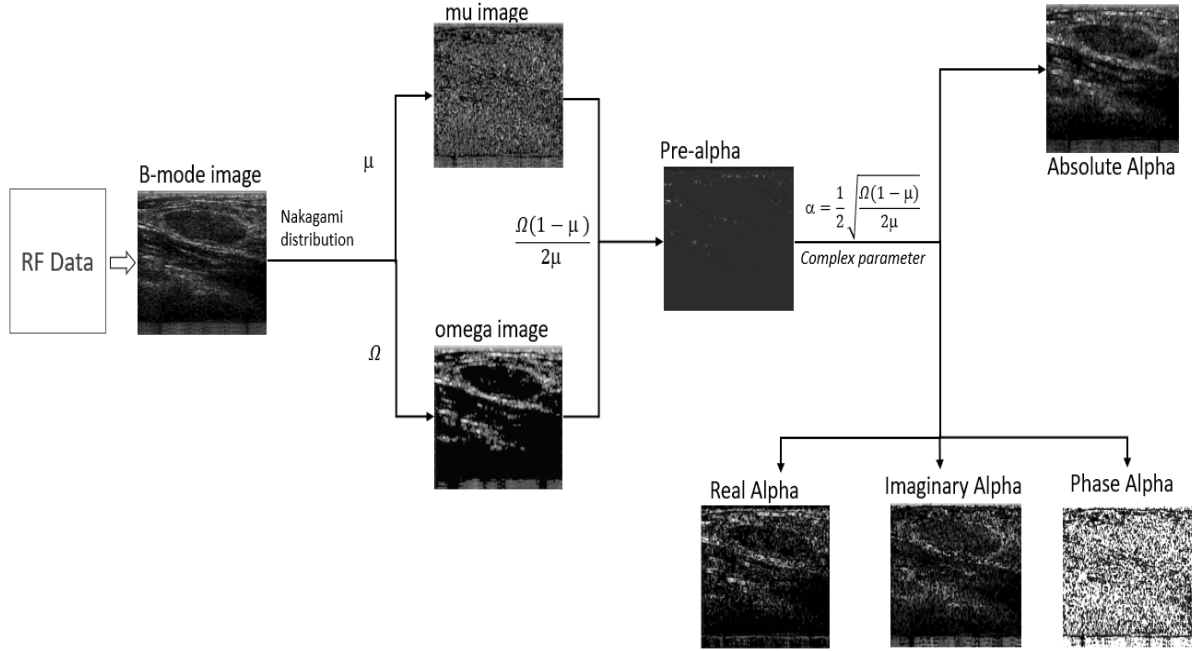

Fig. 2: Overview of the 7 types of Nakagami Parametric Images

**Feature Extraction**

All feature processing software have been built in MATLAB$^{TM}$ (The MathWorks, Inc., Natick, MA). 3 types of features have been extracted from each Nakagami parametric image:

1) Morphometric Features

2) Elemental Features

3) Hybrid Features

*Morphometric Features*

These features describe the shape or boundary of the lesion. They have been calculated based on the lesion boundaries traced on the Nakagami images. These features are described below:

a) Aspect Ratio: Ratio of maximum vertical lesion dimension to maximum horizontal lesion dimension.

$$Aspect\ Ratio = \frac{Maximum\ Vertical\ Lesion\ Dimension}{Maximum\ Horizontal\ Lesion\ Dimension} \quad (17)$$

b) Compactness: Ratio of square root of lesion area and its maximum diameter. It represents the compactness of the shape of the traced boundary.

$$Compactness = \frac{\sqrt{Lesion\ Area}}{Maximum\ Diameter} \quad (18)$$

c) Roundness: Ratio of the lesion area and its maximum diameter squared. It represents the roundness of the shape of the traced boundary.

$$Roundness = \frac{Lesion\ Area}{(Maximum\ Diameter)^2} \quad (19)$$

d) Convexity: Ratio between convex perimeter and actual lesion perimeter which quantifies the border property. This parameter is very sensitive to lesion spiculation and thus it is adept at quantifying border property.

$$Convexity = \frac{Convex\ Perimeter}{Lesion\ Perimeter} \quad (20)$$

e) Form Factor: Ratio of the lesion area to the square of lesion perimeter.

$$Form\ Factor = \frac{Lesion\ Area}{(Perimeter)^2} \quad (21)$$

f) Solidity: Ratio between lesion area and convex area. It is derived as,

$$Solidity = \frac{Lesion\ Area}{Convex\ Area} \quad (22)$$

g) Fractal Dimension for morphometric feature analysis: A fractal dimension is a ratio providing a statistical index of complexity comparing how details in a pattern (strictly speaking, a fractal pattern) change with the scale at which it is measured. It has also been characterized as a measure of the space-filling capacity of a pattern that describes how a fractal scales differently from the space it is embedded in. The fractal dimension of a closed contour is also an indicator of boundary irregularity or roughness (Alam et al., 2011). Fractal dimension has correlation with the quantity roughness, and as spiculation increases fractal dimension increases. We calculated fractal dimensions in three different ways for analyzing the boundary roughness:

- Kolmogorov Fractal Dimension: Computes the slope of the line that plots the number of grid squares through which the lesion boundary passes versus the grid size on log-log axes.

- Minkowski Fractal Dimension: Computes slope of the line that plots the area swept out by circles versus their diameter on log-log axes.

- Hausdorff Fractal Dimension: The Hausdorff dimension is an integer for sets of points that define a smooth shape or a shape that has a small number of corners, and for irregular shapes it becomes non-integer. Here, we used the perimeter to calculate this fractal dimension.

*Elemental Features*

Elemental features are related to the quantitative measures of the parametric images. One of the elemental features, Co-occurrence Contrast, has been extracted from the entire parametric image. This is defined as follows:

- Co-occurrence Contrast: Contrast for the co-occurrence matrix of an image to estimate texture of the image.

The remaining features have been extracted from different regions-of-interest (ROI) of Nakagami parametric images. Nine different ROIs were considered for parameter extraction in this study: (1) left-anterior, (2) left-lateral, (3) left-posterior, (4) tumor-anterior, (5) tumor, (6) tumor-posterior, (7) right-anterior, (8) right-lateral and (9) right-posterior. These are illustrated in Figure 3.

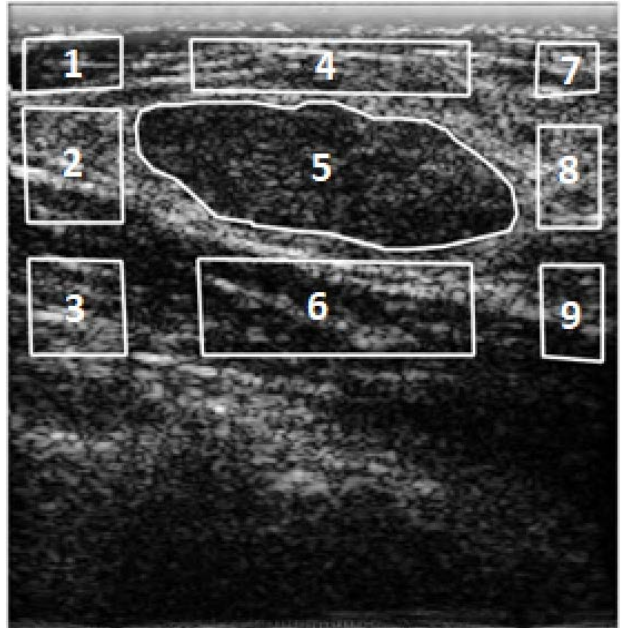

Fig. 3: Nakagami mu image with analysis region traces superimposed: (1): left-anterior, (2): left-lateral, (3): left-posterior, (4): tumor-anterior, (5): tumor, (6): tumor-posterior, (7): right-anterior, (8): right-lateral, and (9): right-posterior

These features are described below:

**a)** Echogenicity: Mean of the pixel intensity values within the lesion, and it is represented by $\mu_L$. For this feature no attenuation correction is necessary.

**b)** Heterogeneity: Standard deviation of pixel intensity values within the lesion and it is represented by $\sigma_L$.

**c)** FNPA: Four-neighborhood pixel algorithm, that computes 4-neighborhood-pixels texture of any parameter within the lesion area (Alam et al., 2011). FNPA for an image of size $m \times n$ with pixel values $x(k,l)$ is given by,

$$FP2 = FP1/\mu \tag{23}$$

where,

$$FP1 = \sum_{l=1}^{n}\sum_{k=1}^{m} \frac{1}{4}[|x(k,l) - x(k-1,l)| + |x(k,l) - x(k,l-1)| + |x(k,l) - x(k+1,l)| + |x(k,l) - x(k,l+1)|] \tag{24}$$

And, $\mu$ is the mean value of $FP_1$.

**d)** Hurst Coefficient Fractal Dimension: The texture of the lesion in an image can be defined using this feature, that is, a Hurst coefficient is a fractal dimension coefficient that characterizes the surface roughness within the lesion. It uses $7 \times 7$ sub-images for the computation.

**e)** Shadow-Normal: Difference between mean Nakagami parameter values in comparable shadowed and non-shadowed regions posterior to the lesion (normalized by lesion thickness). The average of the variations between left-lateral and left-posterior, and right-lateral and right-posterior is compared to the difference between tumor and tumor-posterior. It is defined as follows:

$$Shadow = M_{pl} - \frac{1}{2}(M_{pnr} + M_{pnl}) \qquad (25)$$

where, $M_{pl}$ is the mean Nakagami parameter posterior to lesion, $M_{pnr}$ and $M_{pnl}$ are the mean Nakagami parameters in normal tissue right and left lateral posterior to lesion respectively (Alam et al., 2011).

**f)** Relative-Absorption: Compounded feature and that is given by:

$$Relative\ Absorption = \frac{1}{d1}(M_{pn} - M_{an}) - \frac{1}{d2}(M_{pl} - M_{al}) \qquad (26)$$

where, $M_{al}$ is the mean Nakagami parameter inside the lesion, $M_{pl}$ is the mean Nakagami parameter posterior to the lesion, $M_{an}$ is the mean Nakagami parameter of the normal tissue next to the lesion, $M_{pn}$ is the mean Nakagami parameter in normal tissue lateral posterior to the lesion, $d2$ is the distance between $M_{pl}$ and $M_{al}$ centroids, and $d1$ is the distance between $M_{pn}$ and $M_{an}$ centroids (Alam et al., 2011).

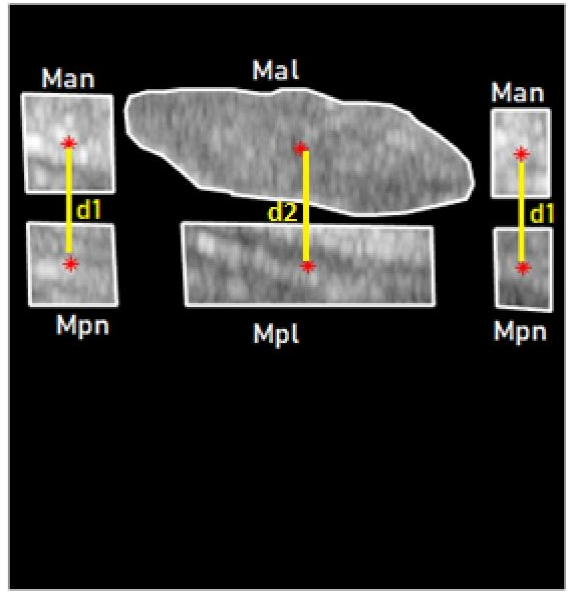

Fig. 4: Relative absorption

*Hybrid Features*

These features are a combination of both morphometric and elemental features. They are calculated using both the lesion contour and elemental features. To eliminate dependence on contour length and Nakagami parameter values, normalization is necessary. These features are described below:

a) Margin definition Area: Ratio of the gradient around the lesion boundary of one pixel width to the area of the boundary region.
  b) Margin definition Gradient: Sum of the magnitudes of the lesion contour's gradients normalized by the sum of the pixel values inside the lesion contour.

For each patient 7 Nakagami parametric images have been generated, from which 7 Elemental and 2 Hybrid features can be obtained for each image. Alongside these features, 9 additional morphometric features are generated for each patient. Thus, in total, 72 features were extracted from each patient scan.

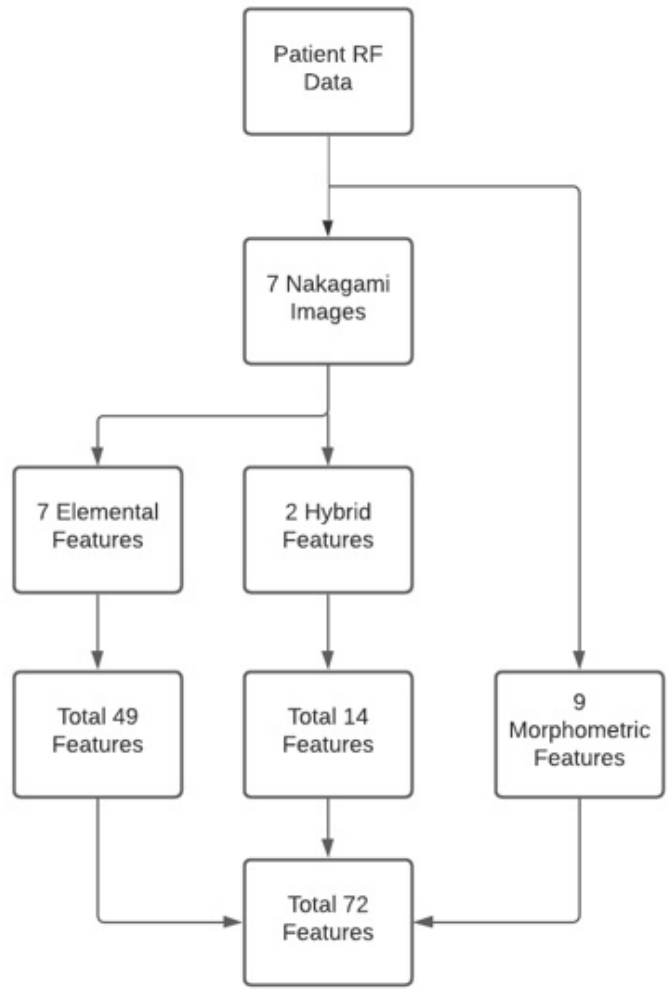

Fig. 5: Breakdown of all the features used in this study

**RFE-CV Feature Reduction**

A dataset with a large number of features is not suitable for machine learning classifiers. Feature Selection (FS) provides a solution to this problem. It encompasses selecting a subset of relevant features from the input feature set that are the most significant in terms of the target concept. The selected subset of features has the ability to improve learning accuracy and reduce training time (Langley, 1995, 1994; Zhao et al., 2010) by selecting only those features which have the most predictive power and removing irrelevant and redundant features (John et al., 1994). Furthermore, reducing the dimensionality of data through feature selection improves the generalization of the developed model by minimizing the risk of overfitting, and hence results in better predictability (Gutlein et al., 2009; Piramuthu, 2004).

In this study, we have utilized Recursive Feature Elimination with Cross Validation (RFE-CV), a wrapper-type feature selection algorithm, to select the subset of most relevant features for our dataset. It is based on the Recursive Feature Elimination (RFE) algorithm, which is very suitable for feature selection of datasets with a small sample size (Guyon et al., 2002). RFE works by eliminating the least important features whose exclusion would have the least effect on the chosen performance metric, for our case it was accuracy. It starts by searching for a subset of features taking the whole training dataset as input and discards less important features until a desired number of features remain. To accomplish this, RFE uses different machine learning models like Decision Tree, Random Forest and Logistic Regression at its core, to compute the redundancy of a feature and rank features by importance. With each iteration the model is refitted with fewer and fewer features and performance is evaluated in terms of the accuracy. For accuracy prediction, the RFE can fit a classifier model predefined by the user to the selected subset of features. In our study, the algorithm, *Multilayer Perceptron (MLP)* led us towards obtaining the best combination of the features for classification with *Linear SVM*.

The main advantage of RFE is flexibility by proper tuning of its hyperparameters, which can be tuned to achieve desired model performance. The main drawback of RFE is that the minimum number of features required for the optimum result has to be predefined as an input. But it is impossible to determine this minimum number and an input has to be given based on speculation which may not actually lead towards the optimum result. To avoid this problem, RFE-CV uses cross validation to determine the minimum number of features for the most optimum result, and no feature number has to be specified as input. RFE-CV scores different subsets of features based on accuracy. For every subset of different sizes, it keeps track of the mean values of the scores obtained from all the test sets while performing cross validation. From these subsets of different sizes, the one with the best mean score is chosen as the best combination of minimum number of features for the optimum result. So, RFE-CV is more accurate than RFE as it makes the feature selection process automated, and leaves no subset untested by trying out every possible feature combination.

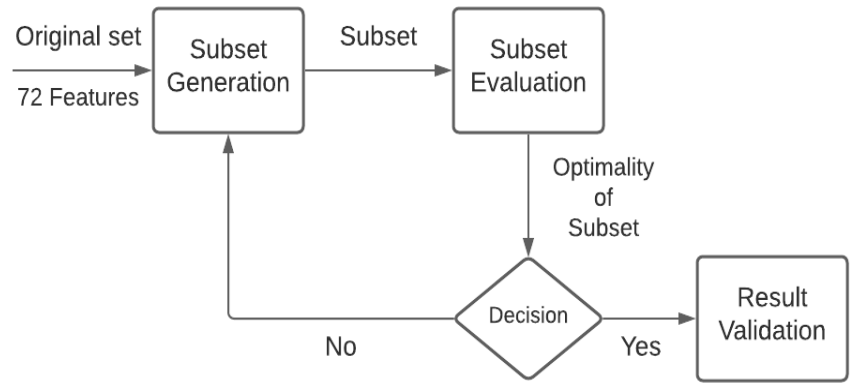

Fig. 6: RFE-CV Feature Selection Algorithm

## RESULTS

Three sets of data were prepared by extracting features from parametric images formed using three windows of different size: 0.1875mm, 0.45mm and 0.75mm. Feature selection using RFE-CV generated a subset of the same five features for data generated using all three window sizes. These five features are, *Aspect Ratio* (morphometric), *Convexity* (morphometric), *Mu Co-occurrence contrast* (Elemental), *Mu Hurst Coefficient* (Elemental) and *Phase Echogenicity* (Elemental). These five features from all three sets of data were used for classification, using *Linear SVM classifier* with *5-fold cross-validation* (from MATLAB Classification Learner Application). An empirical analysis among the three different window sizes (Table 1) clearly delineated an improvement in performance metrics: AUC and Accuracy, with increasing window size, and a conclusion could be drawn that the dataset of five features for images generated using a window size 0.75mm yields the best Area Under the ROC (AUC) value of 0.9712, which is shown in Figure 7.

| Window Size | AUC | Accuracy |
|---|---|---|
| 0.1875mm | 0.87 | 86.9% |
| 0.45mm | 0.93 | 87.7% |
| 0.75mm | 0.97 | 89.2% |

Table 1: Empirical Analysis of Three Window Sizes.

After the selection of the best window size i.e., 0.75mm, the individual performance of the 5 selected features derived from parametric images generated using this window were analyzed. The mean and standard deviation values of these 5 features are provided in Table 2. The distribution of feature values for benign and malignant cases for each of the 5 features are depicted in Figure 8.

| Feature | Benign Cases | Malignant Cases |
|---|---|---|
| Aspect Ratio | $0.634 \pm 0.207$ | $0.887 \pm 0.222$ |
| Convexity | $0.993 \pm 0.011$ | $0.961 \pm 0.029$ |
| Phase Echogenicity | $0.909 \pm 0.123$ | $1.005 \pm 0.086$ |
| Mu Co-occurrence Contrast | $3.232 \pm 2.843$ | $3.279 \pm 2.197$ |
| Mu Hurst Coefficient | $0.5626 \pm 0.0470$ | $0.538 \pm 0.030$ |

Table 2: Mean and standard deviation values of the 5 selected features for a window of 0.75mm.

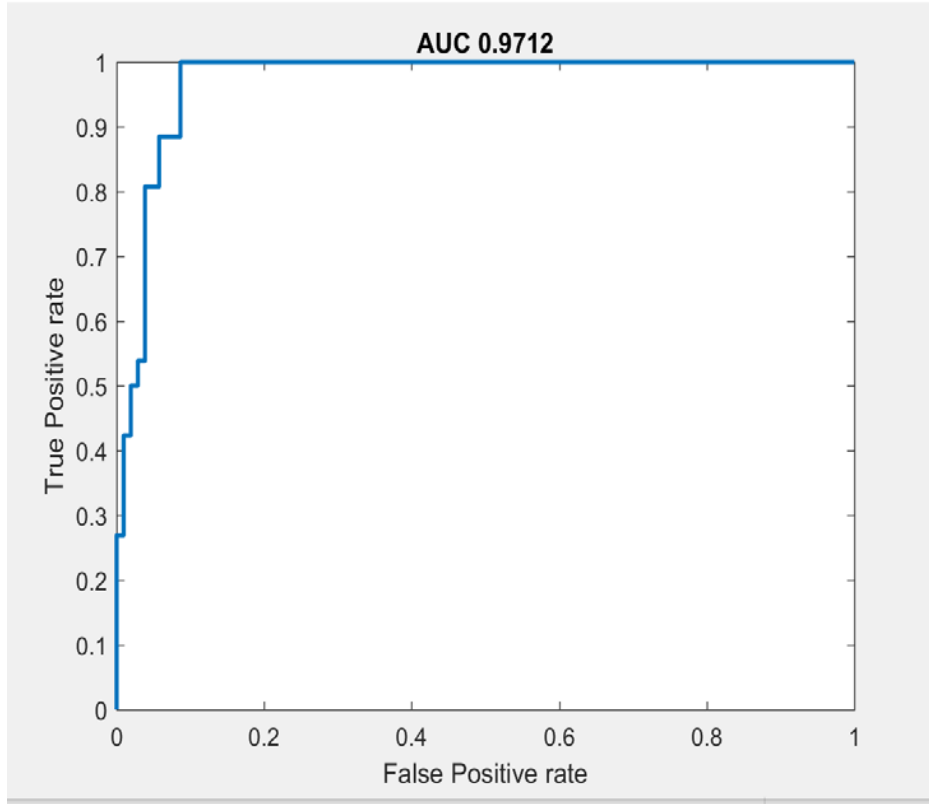

Fig. 7: Area under the ROC Curve for the Dataset of Window Size of 0.75mm

Correlation coefficients between the features were also calculated and all the values were found to be below 0.5. Table 3 depicts correlation among the features.

Table 4 denotes the performance parameters obtained using the 5 selected features extracted from parametric images formed using a window size of 0.75mm. A Classification accuracy of 89.2% was obtained using the selected features, along with a Sensitivity of 0.577 and a Specificity of 0.971.

To classify all the malignant cases accurately (i.e., to make False Negatives, FN=0) and to maintain a reasonable value of the cases requiring further biopsy (i.e., False Positives), hyper parameter tuning was conducted for the threshold value of AUC, keeping the AUC constant to 0.9712. While performing the 5-fold cross validation with *Linear SVM Classifier*, each data from every 5-test dataset was assigned a unique threshold value on which they were later classified. All these threshold values were obtained using the performance curve function from MATLAB. These threshold values ranged from a maximum to a minimum value and any specific value could be chosen as the divider, above which all the data were assigned a certain class, and all the data below the opposite class. While the MATLAB classifier application chose a default divider threshold value and deduced an output based on it, our attempt was to perform classification choosing different threshold values and using that to determine the output. Our goal was to obtain the most optimal threshold in terms of Sensitivity, Specificity as well as Accuracy. Table 5 denotes the accuracy scores and other quantities at different decision thresholds.

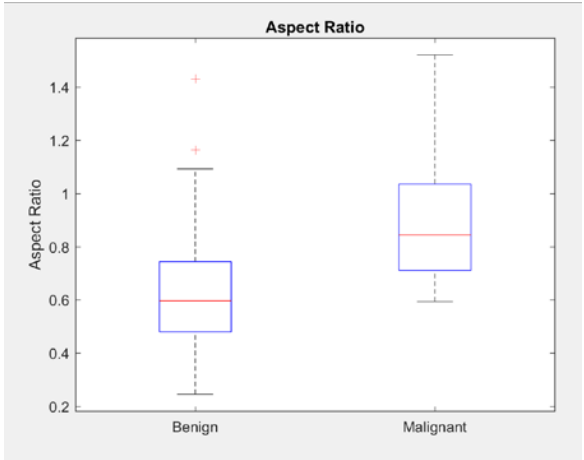
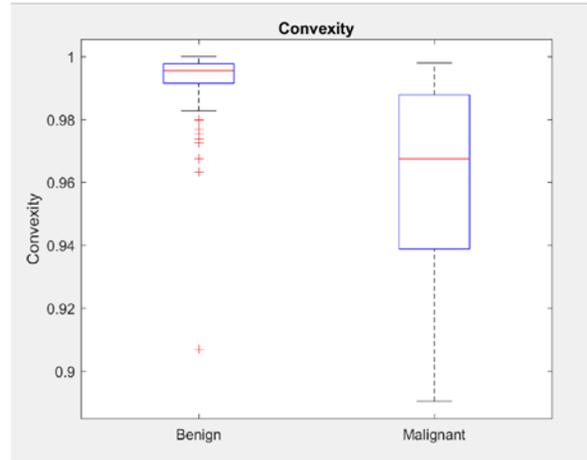

(a)           (b)

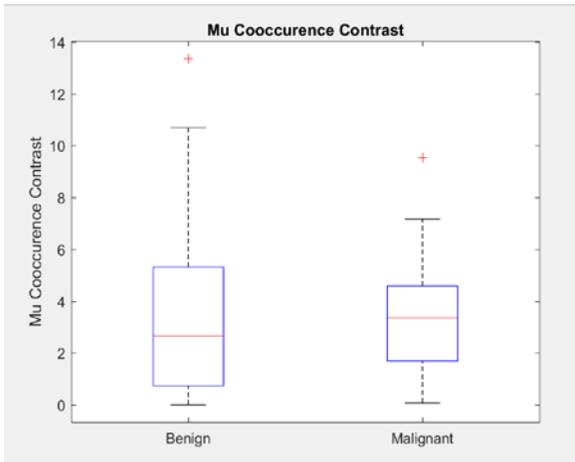
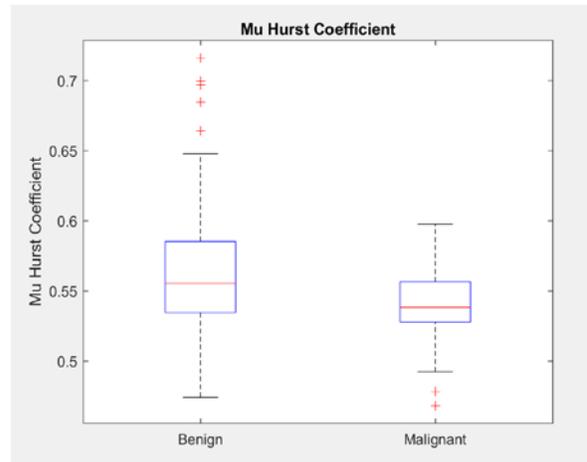

(c)           (d)

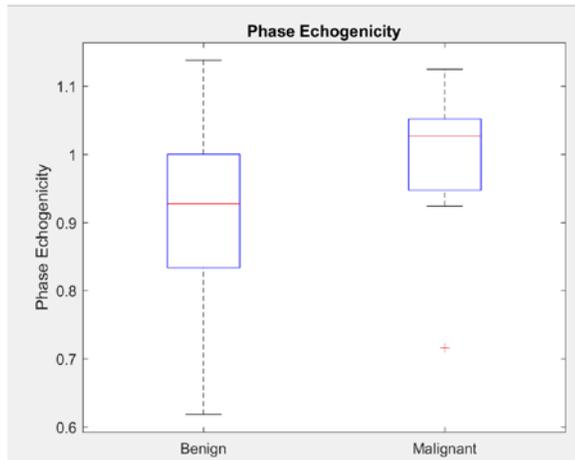

(e)

Fig. 8: Boxplot of the 5 selected features for 0.75mm window size: (a) Aspect Ratio. (b) Convexity. (c) Mu Co-occurrence Contrast. (d) Mu Hurst Coefficient Fractal Dimension. (e) Phase Echogenicity.

| Correlation Coefficients | Aspect Ratio | Convexity | Phase Echogenicity | Mu Co-occurrence Contrast | Mu Hurst Coefficient |
|---|---|---|---|---|---|
| Aspect Ratio | 1.00 | -0.312 | 0.263 | 0.062 | -0.287 |
| Convexity | -0.312 | 1.00 | -0.102 | -0.013 | 0.123 |
| Phase Echogenicity | 0.263 | -0.102 | 1.00 | 0.496 | 0.068 |
| Mu Co-occurrence Contrast | 0.062 | -0.013 | 0.496 | 1.00 | -0.044 |
| Mu Hurst Coefficient | -0.287 | 0.123 | 0.068 | -0.044 | 1.00 |

Table 3. Correlation among the Selected Features for Window Size of 0.75mm.

| False Negative | True Negative | False Positive | True positive | Sensitivity | Specificity | Classification Accuracy |
|---|---|---|---|---|---|---|
| 11 | 101 | 3 | 15 | 0.577 | 0.971 | 89.2% |

Table 4. Performance Parameters before the Tuning of Decision Threshold.

| Threshold | False Negative | True Negative | False Positive | True positive | Sensitivity | Specificity | Classification Accuracy |
|---|---|---|---|---|---|---|---|
| 0.888330 | 0 | 87 | 17 | 26 | 1 | 0.837 | 86.92% |
| 0.849946 | 0 | 89 | 15 | 26 | 1 | 0.856 | 88.46% |
| 0.726128 | 0 | 95 | 9 | 26 | 1 | 0.913 | 93.08% |
| 0.578317 | 3 | 97 | 7 | 23 | 0.885 | 0.933 | 92.30% |
| 0.019403 | 11 | 100 | 4 | 15 | 0.577 | 0.962 | 88.46% |

Table 5. Tuning of Decision Threshold.

From Table 5, it can be determined that a threshold value of 0.726128 provides the optimal circumstances, yielding an accuracy score of 93.08% along with a Sensitivity value of 1 and a Specificity value of 0.913. Furthermore, a threshold of 0.726128 provides an FPR (False Positive Rate) of 8.65%, which is significantly small, along with a FNR (False Negative Rate) of 0%.

**DISCUSSION**

Ultrasound instruments are becoming more effective in distinguishing between cancerous and non-cancerous solid breast tumors visually, and this has moved to a point where many radiologists recommend periodic follow-ups based on breast ultrasound findings, without performing a biopsy. We studied the classification performance of multiple features to determine whether such an approach might play a role in reducing misdiagnoses or unnecessary biopsies. The results obtained from this study indicate that our approach has the potential to distinguish between benign and malignant breast lesions with a very high accuracy. Furthermore, upon tuning the decision threshold, not only were we able to reduce the number of False Negatives to 0, the optimum threshold value resulted in a very minimal False Positive Rate (FPR) of 8.65%. The final output of our proposed system yielded a classification accuracy of 93.08%, along with a Sensitivity of 1 and a Specificity of 0.913. Thus, we believe that our proposed system has the potential to assist clinicians in diagnosis, as well as significantly reduce the number of unnecessary biopsies.

Our system began with a total of 72 features extracted from 7 different types of Nakagami parametric images. A mixture of morphometric features, elemental features and hybrid features were extracted from each type of parametric image. Feature selection was then performed to identify the subset of features that are the most effective in distinguishing between benign and malignant masses. This final subset constituted 5 features: Aspect Ratio, Convexity, Mu Co-occurrence contrast, Mu Hurst Coefficient and Phase Echogenicity.

The Aspect Ratio and Convexity are morphometric features that remain the same across all 7 images. The remaining three features selected by our feature selection algorithm are elemental. The elemental features examined in this study may be affected by breast composition (e.g., fatty, fibrous or glandular breast types), as this varies between different people, and are likely to influence RF echoes. In this case, the contralateral breast, which is presumed to be healthy, may be used to provide a baseline to compare with the diseased breast (Kaplan, 2001). Two out of the three elemental features are from the $m$ parametric image. The Nakagami $m$ parameter has been found to be effective towards the classification of benign and malignant masses in previous studies (Klimonda et al., 2019; Shankar et al., 2003, 2002, 2001). In our study, both features that were selected for classification from the $m$ Parametric images are related to image texture. Thus, texture seems to be a good indicator for characterizing benign and malignant masses, and future studies may analyze other textural features derived from the $m$ parametric image. The third elemental feature is derived from the $\alpha$ - phase image, or related to the phase of effective cross-section of scatterers. As discussed previously, the complex parameter $\alpha$ is dependent on the number density of scatterers, the signal-to-noise ratio of the cross section as well as attenuation. Hence, this parameter has the potential to be very effective in tissue characterization, and the significance of the phase of this parameter may be explored in greater detail in future studies.

Hybrid features were not found to be effective in distinguishing between tumor types in this study. The algorithm we used for feature selection didn't consider the hybrid features as a part of the best subset of features for classification, although this may differ with other FS algorithms.

In Figure 8, boxplot pairs corresponding to values of the selected 5 features determined for benign and malignant lesions are analyzed. The morphometric features seem to be the most effective in distinguishing between lesion types. From Figure 8(b) it is seen that the feature Convexity has the highest distinguishing ability, followed closely by Aspect Ratio. Overlapping is observed for the elemental features, although the median values are clearly separable between benign and malignant cases. Correlation between the selected features is analyzed in Table 3, and no features were found to be significantly correlated.

The dataset utilized for this study contained an imbalance between positive and negative classes (104 positive, 26 negative). Although we achieved very satisfactory performance, completely eliminating False Negatives and reducing the False Positive Rate to 8.65%, future studies are recommended to utilize a balanced dataset to avoid classifier biasing.

**CONCLUSION**

In this study, we propose a framework to classify cancerous and non-cancerous tumors of the breast using a combination of comprehensive statistical features and machine learning. Our results indicate that such a technique can play a significant role in the diagnostic process associated with detection of breast cancer, as we were able to distinguish between tumor types with a high level of accuracy, as well as provide a very low percentage of false positives. The scope of this study can be further enhanced by incorporating statistical distributions such as the homodyne-K distribution. The current process remains semi-automated because of the need to manually segment the lesion, however, automated boundary detection using sophisticated image-segmentation methods may be implemented in the future to make the process automated. Furthermore, a larger dataset may be used to validate the results obtained in this study. Deep learning models such as the Convolutional Neural Network may also be incorporated to extend the scope of this study further.